\documentclass{article}
\usepackage{spconf,amsmath,graphicx}
\usepackage{booktabs}
\usepackage{multicol}
\usepackage{multirow}
\usepackage{graphicx}
\usepackage{amssymb}
\usepackage{bm}
\usepackage{url}
\usepackage{cite}
\usepackage[colorlinks,linkcolor=black,anchorcolor=black,citecolor=black,urlcolor=black]{hyperref}

\title{Speech reconstruction from silent tongue and lip articulation by pseudo target generation and domain adversarial training}
\name{Rui-Chen Zheng, Yang Ai*, Zhen-Hua Ling\thanks{* Corresponding Author. This work was partially supported by the Fundamental Research Funds for the Central Universities.}}
\address{National Engineering Research Center of Speech and Language Information Processing, \\University of Science and Technology of China, Hefei, P. R. China\\
{\small \tt zhengruichen@mail.ustc.edu.cn,  \{yangai, zhling\}@ustc.edu.cn}}
\begin{document}
\ninept
\maketitle
\begin{abstract}
This paper studies the task of speech reconstruction from ultrasound tongue images and optical lip videos recorded in a silent speaking mode, where people only activate their intra-oral and extra-oral articulators without producing sound. This task falls under the umbrella of articulatory-to-acoustic conversion, and may also be refered to as a silent speech interface. We propose to employ a method built on pseudo target generation and domain adversarial training with an iterative training strategy to improve the intelligibility and naturalness of the speech recovered from silent tongue and lip articulation. Experiments show that our proposed method significantly improves the intelligibility and naturalness of the reconstructed speech in silent speaking mode compared to the baseline TaLNet model. When using an automatic speech recognition (ASR) model to measure intelligibility, the word error rate (WER) of our proposed method decreases by over 15\% compared to the baseline. In addition, our proposed method also outperforms the baseline on the intelligibility of the speech reconstructed in vocalized articulating mode, reducing the WER by approximately 10\%. 
\end{abstract}

\vspace{-1mm}
\section{Introduction}
\label{sec:intro}
\vspace{-1mm}

The speech production process involves the coordination of a series of vocal organs, such as the tongue, jaw, velum, and lips \cite{denes1993speech}. Therefore, articulatory features and acoustic features are intrinsically linked \cite{liu2016articulatory}. Articulatory-to-acoustic conversion is a task derived from the above theory, aiming to synthesize acoustic features directly from articulatory input \cite{kello2004neural, csapo2020ultrasound}. It can also be referred to as silent speech interface (SSI), which relies on non-acoustic signals generated by the speakers during the speech production process to enable communication although the regular verbal communication is impossible \cite{denby2010silent, schultz2017biosignal, gonzalez2020silent}. SSI is of tremendous research significance because it can help restore speech communication for users with dysphonia and assist communication when speech is not available or desirable.

People speak in different modes in different scenarios. Under most circumstances, speakers adopt the standard vocalized speaking mode, which means their larynx and lungs function as expected. Over the past few years, there has been a great deal of work on SSI in the vocalized speaking mode. An early study adopted an hidden Markov model (HMM) based statistical model and a unit selection algorithm to predict the acoustic features corresponding to the features from vocalized tongue and lip articulation \cite{hueber2010development}. After the rapid development of deep learning, deep neural networks (DNNs) and convolutional neural networks (CNNs) have been developed for speech reconstruction based on vocalized tongue or lip movement recordings \cite{csapo2017dnn, akbari2018lip2audspec, kumar2019lipper}. Besides, there was a study using bidirectional long short-term memory (BLSTM) based recurrent neural networks (RNN) to convert recorded electromagnetic midsagittal arthrography (EMA) measurements into spectral and excitation features, and finally restore speech waveforms \cite{liu2018articulatory}. Recently, a model based on encoder-decoder architecture named TaLNet has been proposed to reconstruct speech from both the vocalized tongue ultrasound and lip video, leveraging transfer learning from text-to-speech (TTS) models and achieving impressive results \cite{zhang2021talnet}.

In some situations where silence is required, or for some laryngectomy patients, speakers tend to take the silent speaking mode, which means during articulation, speakers only activate their oral and nasal articulators but suppress their laryngeal activity, and consequently, no sound is produced as output. Previous works pay little attention to the performance of speech reconstructed in silent speaking mode. To reconstruct speech from silent articulation faces the following two challenges.
First, a large number of studies have proven that there exist discrepancy between vocalized and silent articulation. The lack of intra-oral pressure in silent speaking mode can cause speakers to produce incomplete or reduced articulators movements \cite{ribeiro2021silent}. The articulatory movements during silent articulation last longer \cite{zhang2021talnet, ribeiro2021silent, dromey2017effects}, and show a decrease in peak velocity and an increase in the number of articulatory sub-movements \cite{teplansky2019tongue}. Moreover, phonemes produced in the silent mode exhibit less obvious tongue movement patterns, manifested by a reduced spatial area of articulation distinctiveness \cite{teplansky2019tongue, teplansky2020tongue}. These findings reveal that silent and vocalized articulations belong to two different domains. Hence, the models trained on vocalized data cannot be applied to the silent mode directly. Second, the silent speaking mode produces no 
speech signals. 
Therefore, the model with silent articulation as input cannot be trained using the same supervised paradigm as that in the vocalized mode. 

To address the above challenges of speech reconstruction from silent tongue and lip articulation, this paper proposes the following approaches. (1) To address the issue of no corresponding natural speech output for training in silent speaking mode, we use dynamic time warping (DTW) \cite{muller2007dynamic} to generate pseudo targets for unlabeled silent articulating data. (2) Since the articulation state in the silent speaking mode is more uncertain than that in the vocalized mode, we combine the vocalized and silent data, and introduce domain adversarial training to learn robust representations invariant in both vocalized and silent domains. (3) An iterative training strategy is designed, which iteratively conducts the first two steps to further boost model performance. 
Experimental results on the Tongue and Lip (TaL) dataset \cite{ribeiro2021tal} show that our proposed method effectively improves the intelligibility and naturalness of the reconstructed speech in the silent speaking mode while achieving some degree of advancement in the vocalized mode.

\vspace{-1mm} 
\section{Related Work}
\label{sec:format}
\vspace{-1mm}

Our proposed method is built based on TaLNet \cite{zhang2021talnet}, the state-of-the-art model for speech reconstruction with vocalized tongue and lip articulation as input on TaL dataset \cite{ribeiro2021tal}. It has an encoder-decoder architecture. The encoder of TaLNet first encodes the input tongue images and lip videos into articulatory representations, and the decoder then decodes them into acoustic features. The acoustic features are ultimately fed into a well-trained neural vocoder to generate the final speech waveforms. 

The encoder of TaLNet includes two parallel sub-encoders dedicated to processing ultrasound tongue images and optical lip video, respectively. Specifically, each sub-encoder consists of stacked 3D CNNs. A single visual vector for each input frame is generated through the sub-encoder. Finally, the vectors encoded from tongues and lips seperately are fused through concatenation and linear projection to produce the articulatory representations. 

The decoder of TaLNet is migrated from the Tacotron2-based \cite{shen2018natural} TTS acoustic model, except that a forced-attention mechanism substitutes for the standard soft-attention. Each frame of acoustic features is predicted by the decoder in an autoregressive manner. The autoregressive input is first processed by a pre-processing network and then sent to a two-layer long short-term memory (LSTM) network.  A CNN-based post-processing network is employed after the initial decoder output to refine the acoustic feature prediction.

For training TaLNet, a multi-speaker Tacotron2 model is first built on a multi-speaker TTS corpus. Then its decoder is transferred as a TaLNet decoder, and meanwhile, the soft attention is replaced with the forced attention. The transferred decoder is then jointly trained with the encoder of TaLNet. For more details of TaLNet, we suggest readers refer to its original paper \cite{zhang2021talnet}.

\vspace{-1mm}
\section{Proposed Method}
\label{sec:pagestyle}
\begin{figure}[htbp]
        \setlength{\abovecaptionskip}{0.cm}
        \setlength{\belowcaptionskip}{-1cm}
	\centering
	\includegraphics[width=3.20in]{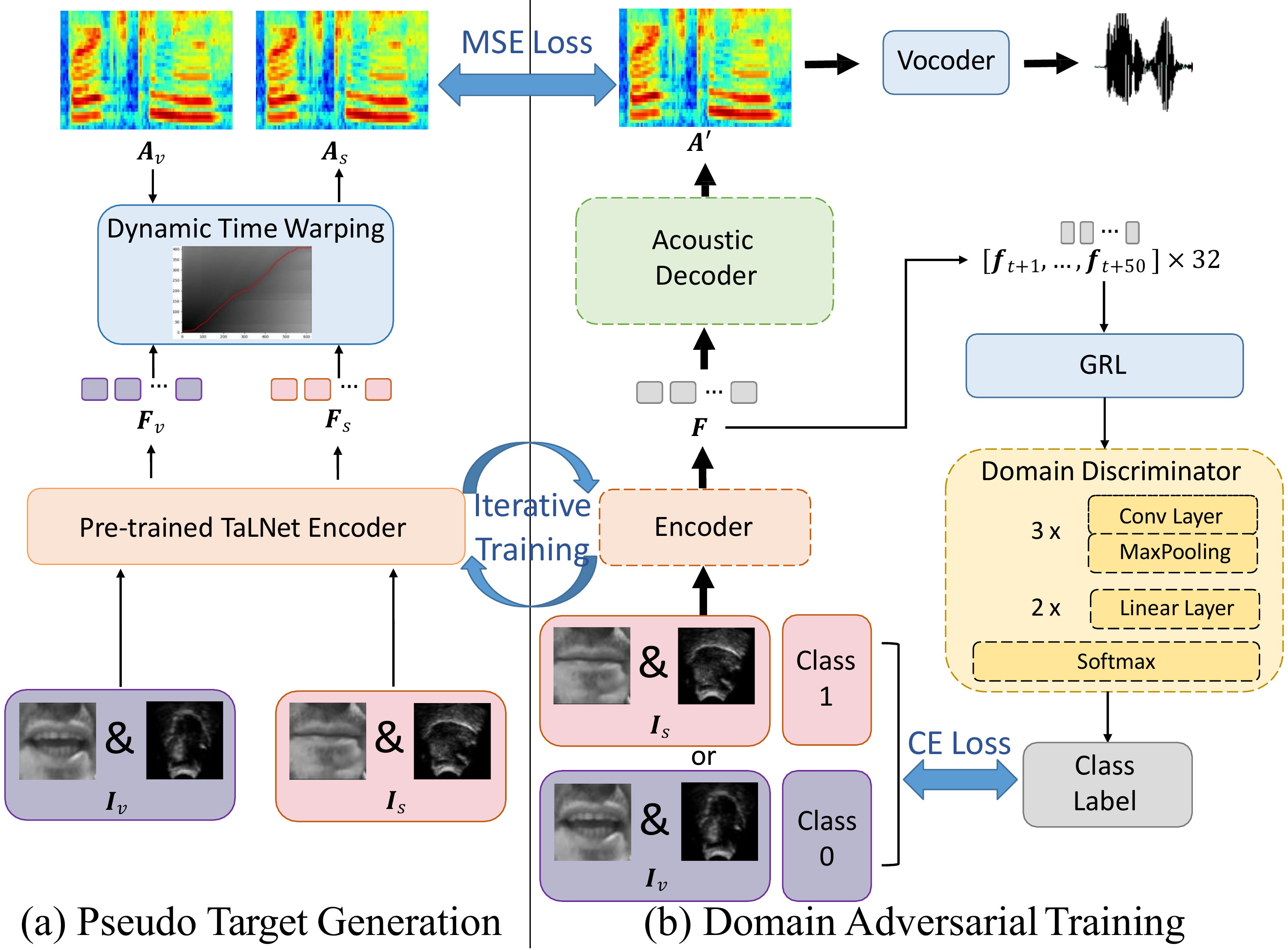}
	\caption{Details of the proposed method. The dashed box indicates that the parameters of this part are updated during training. The process of inference is shown by the bold black arrows. Symbols appearing in the figure are defined in Section 3.1.}
	\label{TaLNet-DA}
        \vspace{-1mm}
\end{figure}

\vspace{-1mm}
This paper proposes to employ pseudo target generation, as well as domain adversarial training, to address the two major challenges of SSI tasks in the silent speaking mode. An iterative training strategy is also designed to further boost model performance. In our proposed model, the structures of both the encoder and the decoder are consistent with the ones in TaLNet \cite{zhang2021talnet}. Details are illustrated in Fig.~\ref{TaLNet-DA} and will be introduced in this section. 

\vspace{-1mm}
\vspace{-1mm}

\vspace{-1mm}
\subsection{Pseudo Target Generation}
\label{subsec:ptg}
\vspace{-1mm}
In the silent speaking mode, there is no speech signal generation since the vocal cord does not vibrate, which makes it challenging to apply traditional supervised sequence-to-sequence paradigms. To solve this problem, we propose to generate pseudo acoustic targets for silent articulation so that the same supervised learning paradigm can be employed as in the vocalized mode. The general way to produce pseudo targets for an unlabeled target domain is directly utilizing the prediction results of the model on the target domain. However, the model trained on the source domain is usually unable to perform well on the target domain, leading to pseudo targets of poor quality. We thus bring out a new pseudo target generation method 
making use of the parallel recordings with same linguistic contents between the two modes in the TaL dataset \cite{ribeiro2021tal}.

As shown in Fig.~\ref{TaLNet-DA}(a), a pre-trained TaLNet encoder is first used to encode the silent input $\bm{I}_s$ and its corresponding vocalized input $\bm{I}_v$ with the same linguistic content, and obtain the outputs $\bm{F}_s$ and $\bm{F}_v$. Usually the duation $T'$ of silent input $\bm{I}_s$ is larger than the duration $T$ of vocalized input $\bm{I}_v$ because the articulatory movements during silent articulation always last longer than that in the vocalized articulation. Then we propose to perform dynamic time warping (DTW) \cite{muller2007dynamic} to align the encoder output $\bm{F}_v$ in the source vocalised domain to $\bm{F}_s$ in the target silent domain. From this, an alignment path for aligning acoustic features is obtained. After that, the natural acoustic features $\bm{A}_v$ of the vocalised articulation data are temporally aligned 
along the alignment path obtained by DTW to generate pseudo acoustic features $\bm{A}_s$ on the target silent domain. 
After acquiring pseudo acoustic features for $\bm{A}_s$ silent inputs, the model can be trained in the target silent domain following the supervised training paradigm of TaLNet in the vocalized mode. 

In practice, the DTW algorithm sometimes generates inappropriate alignments, leading to the pseudo targets of poor quality for some training samples. To address this issue, only the samples with reliable pseudo targets are selected for supervised training in our implementation. The cost of DTW alignment is employed as the criterion of selecting reliable samples. Let $D_s$ denote the set of all silent samples. Then, $D_s^{\epsilon}\subset D_s$ is the set of reliable silent samples whose DTW costs are less than a specific threshold $\epsilon$. Thus, the reconstruction loss of using pseudo targets for model training, i.e., the MSE loss in Fig.~\ref{TaLNet-DA}, can be written as
\vspace{-1mm}
\begin{equation}
    \setlength{\abovecaptionskip}{0.cm}
    \setlength{\belowcaptionskip}{-1cm}
    L_{Recon,s} = \mathbb{E}_{D_s^\epsilon}||\bm{A}_s- \bm{A}_s'||_2^2,
\end{equation}
where $\bm{A}_s'$ denotes the acoustic features predicted from silent articulations by the model.

\vspace{-1mm}
\subsection{Domain Adversarial Training}
\label{subsec:dat}
\vspace{-1mm}
Considering the difficulty of collecting articulation data, especially for the silent articulation scenario, our method performs joint training on both silent and vocalized articulation data. For vocalized articulation data, the reconstruction loss can be written as
\vspace{-1mm}
\begin{equation}
        \setlength{\abovecaptionskip}{0.cm}
        \setlength{\belowcaptionskip}{-1cm}
    L_{Recon,v} = \mathbb{E}_{D_v}||\bm{A}_v- \bm{A}_v'||_2^2,
\end{equation}
\vspace{-1mm}
where $\bm{A}_v'$  denotes the acoustic features predicted from vocalized articulations, and $D_v$ means the set of all  vocalized training samples. 
Moreover, as mentioned earlier, there is a domain discrepancy between vocalized and silent articulations. To address this issue, domain adversarial training is introduced into the model to force the encoder to learn more robust articulatory representations invariant in both silent and vocalized domains. The detailed structure is illustrated in Fig.~\ref{TaLNet-DA}(b). Specifically, a domain discriminator is plugged in after the encoder, which accepts the output of the encoder $\bm{F}$ as input to judge whether it comes from the silent domain or the vocalized domain. Due to the different lengths of articulatory representations $\bm{F}$, the outputs of the encoder cannot be directly fed into the domain discriminator. Therefore, 32 segments with a length of 50 frames are randomly selected from the output of the encoder, spliced together, and sent to the domain discriminator for classification. The structure of the domain discriminator consists of a stack of CNN, maxpooling, and linear layers. Finally, a softmax function is used to produce the classification probabilities. 

According to the idea of adversarial training \cite{creswell2018generative}, the discriminator network is trained by minimizing the cross entropy (CE) loss $L_D$ of the discriminator  to enhance its discrimination ability,  while the encoder is expected to maximize $L_D$ in order to generate more confusing representations to cheat the discriminator. Thus, the overall loss function for training the encoder and the acoustic decoder and  can be written as 
\vspace{-1mm}
\begin{equation}
\label{lossoverall}
    \setlength{\abovecaptionskip}{0.cm}
    \setlength{\belowcaptionskip}{-1cm}
    L = L_{Recon,s}+L_{Recon,v}-\lambda L_D,
\end{equation}
where the hyper-parameter $\lambda$ controls the strength of the domain adversarial training. A gradient reversal layer (GRL) \cite{ganin2015unsupervised} is inserted between the encoder and the domain discriminator to achieve the negative of $L_D$ in Eq. (\ref{lossoverall}).

\vspace{-1mm}
\subsection{Iterative Training Strategy}
\vspace{-1mm}
The quality of pseudo targets produced by our proposed method is highly correlated with the appropriateness of the articualtory representations given by the encoder. At the beginning of model training, the pre-trained TaLNet encoder is used for DTW alignment and pseudo target generation. However, the original TaLNet model was trained only with vocalized data, and may give inappropriate articulatory representations for silent data. Therefore, an iterative training strategy is designed to address this issue. To be concrete, after training some epochs, the pseudo targets are generated once again using the encoder updated by the domain adversarial training in Section \ref{subsec:dat}. This process can be repeat iteratively to gradually improve the quality of generated pseudo targets.
Since the domain adversarial training may change the scale of articulatory representations and DTW alignment costs, the threshold $\epsilon$ for selecting reliable silent training samples in Section \ref{subsec:ptg} is also updated correspondingly.

\vspace{-1mm}
\section{Experiments}
\label{sec:typestyle}
\vspace{-1mm}

\vspace{-1mm}
\subsection{Datasets}
\vspace{-1mm}
The Tongue and Lip (TaL) dataset \cite{ribeiro2021tal} was adopted in our experiments. It contains synchronized audio, ultrasound tongue images, and lip videos in both vocalized and silent speaking modes from 81 native English speakers. There are 1212 utterances in the silent speaking mode in the corpus, each with a corresponding vocalized utterance with the same linguistic content. 
Since each speaker has only about 15 utterances in the silent mode, 2 utterances 
were randomly selected from each speaker to form the validation set and the test set respectively, and the rest utterances were used as the training set. 
For domain adversarial training, we did not use all the vocalized utterances in the dataset as \cite{zhang2021talnet}, but only selected the ones with the same linguistic contents as the silent articulation data and combined them into training set. A vocalized test set was also constructed, including one vocalized utterance from each speaker which did not overlap with the vocalized training set of the pre-trained TaLNet and our proposed method.

\vspace{-1mm}
\subsection{Experimental Settings}
\vspace{-1mm}
The TaLNet model built in the previous work \cite{zhang2021talnet} trained on totally 11487 vocalized utterances in TaL dataset was adopted as the baseline model in our experiments.
Considering the limited number of silent utterances for each speaker, we built our proposed model in a speaker-independent way without further finetuning with speaker-dependent data.
For fair comparison, the TaLNet model was also a speaker-independent one without speaker-dependent finetuning.

The encoder of the pre-trained TaLNet model was used for pseudo target generation at the first iteration of iterative training. For domain adversarial training, to suppress the noisy signal from the domain discriminator at the early training stage, we gradually increased the  $\lambda$ in Eq. (\ref{lossoverall}) from 0 using the following schedule  \cite{ganin2015unsupervised}
\vspace{-1mm}
\begin{equation}
        \setlength{\abovecaptionskip}{0.cm}
        \setlength{\belowcaptionskip}{-1cm}
    \lambda = \frac{2}{1+\exp{\{-2 \cdot \frac{current \  epoch}{total \  epochs}\}}}-1.
\end{equation}
In the first 50 training epochs, the encoder's parameters were updated every five batches while the domain discriminator's parameters were updated every single batch. 
In the rest training epochs, the updating frequencies of the encoder and the domain discriminator were switched. A well-trained Parallel WaveGAN (PWG) \cite{yamamoto2020parallel} was used to reconstruct speech waveforms from the predicted mel-spectrograms in our implementation.

Three iterations of iterative training were conducted in our implementation.
The DTW cost threshold $\epsilon$ for selecting reliable training samples were set as 40 for the first iteration and 49 for the following two iterations heuristically according to the means of DTW costs of all training samples.
There were 18 speakers whose DTW costs of all silent utterances were above the threshold in the first iteration.
We suspected that these speakers may have unreliable silent articulations and thus excluded them
from the test set for all silent-mode models in following experiments.


\vspace{-1mm}
\subsection{Experimental Results}
\vspace{-1mm}

Our proposed method was compared with the baseline TaLNet model in both silent (S) and vocalized (V) speaking modes.\footnote{Our demo is available at: \url{ https://zhengrachel.github.io/ImprovedTaLNet-demo/}.} For objective evaluation, mel-cepstral distortion (MCD), short-term objective intelligibility (STOI), and the word error rate (WER) given by a speech recognition engine were used as metrics. Since there was no ground truth speech for the silent utterances, the generated pseudo acoustic features of test utterances were input to the PWG vocoder to generate waveforms, and the waveforms were used as the reference speech of test silent utterances while calculating MCD and STOI. To further evaluate the intelligibility of reconstructed speech, the iFLYTEK speech recognition API\footnote{\url{https://www.xfyun.cn/services/lfasr}} was employed to measure the WERs of different models. For reference, the mean WER of all the vocalized natural recordings in the test set was 4.110\%, and the mean WER of 
the reference speech of test silent utterances mentioned above 
was 10.246\%. Two groups of subjective listening tests were also conducted to measure the naturalness mean opinion scores (MOS) of reconstructed speech in the two modes respectively. In each test, twenty-one native English speakers were recruited on Amazon’s Mechanical Turk\footnote{\url{https://www.mturk.com/}} and were asked to give a 5-point score (1-very poor, 2-poor, 3-fair, 4-good, 5-excellent) for each utterance they listened to. Fifteen utterances generated by each system in each mode were randomly selected for MOS evaluation.

\begin{table}
\renewcommand\arraystretch{0.40}
\setlength{\abovecaptionskip}{0.cm}
\setlength{\belowcaptionskip}{-1cm}
  \centering
    \caption{Objective and subjective evaluation results of speech recovered in silent (S) and vocalized (V) speaking modes. GT represents the vocoder-resynthesized natural speech in the vocalized speaking mode. Best results are highlighted in bold. All results are the means on the test set. ± represents 95\% confidence intervals.}
    \begin{tabular}{c|c|c|c|c|c}
    \toprule
    \toprule
    \multicolumn{1}{l|}{Mode} & Method & \multicolumn{1}{l|}{MCD/dB} &  \multicolumn{1}{l|}{STOI} & \multicolumn{1}{l|}{WER/\%} & \multicolumn{1}{l}{MOS} \\
    \midrule
    \multirow{2}[0]{*}{S} & TaLNet & 4.423 & 0.432 & 59.960 & 2.990±0.123 \\
          & Ours & \textbf{3.935} & \textbf{0.517} & \textbf{43.114} & \textbf{3.330±0.120} \\
    \midrule
    \multirow{3}[1]{*}{V} & TaLNet & 3.421 & 0.666 & 26.890 & 3.421±0.122 \\
          & Ours & \textbf{3.382} & \textbf{0.684} & \textbf{17.309} & 3.672±0.109 \\
          & GT & - & - & - & \textbf{4.161±0.096} \\
    \bottomrule
    \bottomrule
    \end{tabular}%
  \label{eval-results}
  \vspace{-1mm}
\end{table}
\vspace{-1mm}

The evaluation results in the silent mode are illustrated in the first two rows of Table \ref{eval-results}. We can see that the proposed method outperformed the baseline TaLNet model on all objective and subjective metrics, especially achieved a decrease of WER by 15\% and a significant increase of MOS by 0.34 ($p=2.11\times10^{-6}$ in paired t-test), reflecting higher speech intelligibility and naturalness.  

The evaluation results in the vocalized mode are illustrated in the last three rows of Table \ref{eval-results}.
We can see that the proposed method significantly improved the naturalness and intelligibility of the speech recovered from vocalized articulation data using TaLNet. 
The reason can be attributed to that the pseudo labels generated by DTW can be considered as a kind of data augmentation to provide more useful training data. Besides, 
the domain discriminator makes the encoder concentrate on deriving 
domain-invariant articulatory representations, 
which may also improve the generalization ability of the model when dealing with unseen vocalized articulation data.

\vspace{-1mm}
\begin{table}
        \setlength{\abovecaptionskip}{0.cm}
        \setlength{\belowcaptionskip}{-1cm}
	\renewcommand\arraystretch{0.40}
	\centering
	\caption{Objective evaluation results of the proposed method in ablation studies. Here, ITS and DAT stand for ``iterative training strategy" and ``domain adversarial training", respectively. Best results are highlighted in bold.}
	\begin{tabular}{c|c|c|c|l}
		\toprule
		\toprule
		\multicolumn{1}{l|}{Mode} & Method & \multicolumn{1}{l|}{MCD/dB} & \multicolumn{1}{l|}{STOI} & \multicolumn{1}{l}{WER/\%} \\
		\midrule
		\multirow{3}[2]{*}{S} & Ours & \textbf{3.935} & \textbf{0.517} &  \textbf{43.114} \\
		& w/o ITS & 3.971 & 0.512 & 46.016 \\
		& w/o DAT & 4.009 & 0.51  & 51.199 \\
		\midrule
		\multirow{3}[2]{*}{V} & Ours & \textbf{3.382} & \textbf{0.684} &  \textbf{17.309} \\
		& w/o ITS & 3.392 & 0.683 & 18.122 \\
		& w/o DAT & 3.434 & 0.670 & 25.032 \\
		\bottomrule
		\bottomrule
	\end{tabular}%
	\label{ablation}%
 \vspace{-1mm}
\end{table}%
\vspace{-1mm}

\vspace{-1mm}
\subsection{Analysis}
\vspace{-1mm}

\begin{figure}
\setlength{\abovecaptionskip}{0.cm}
\setlength{\belowcaptionskip}{-1cm}
\begin{minipage}[b]{0.5\linewidth}
  \centering
  \centerline{\includegraphics[width=3.4cm]{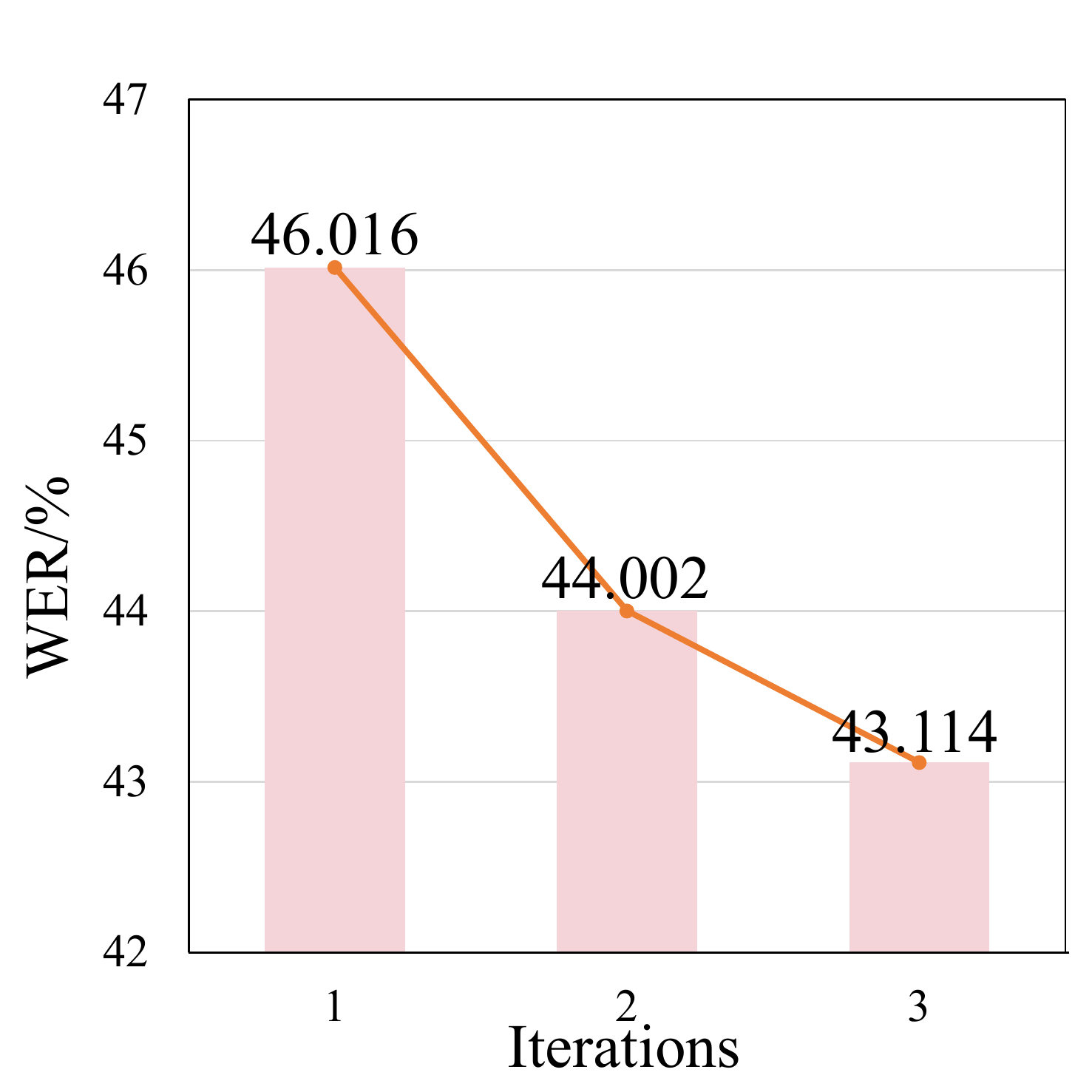}}
  \centerline{(a) Silent Speaking Mode}\medskip
\end{minipage}
\hfill
\begin{minipage}[b]{0.5\linewidth}
  \centering
  \centerline{\includegraphics[width=3.4cm]{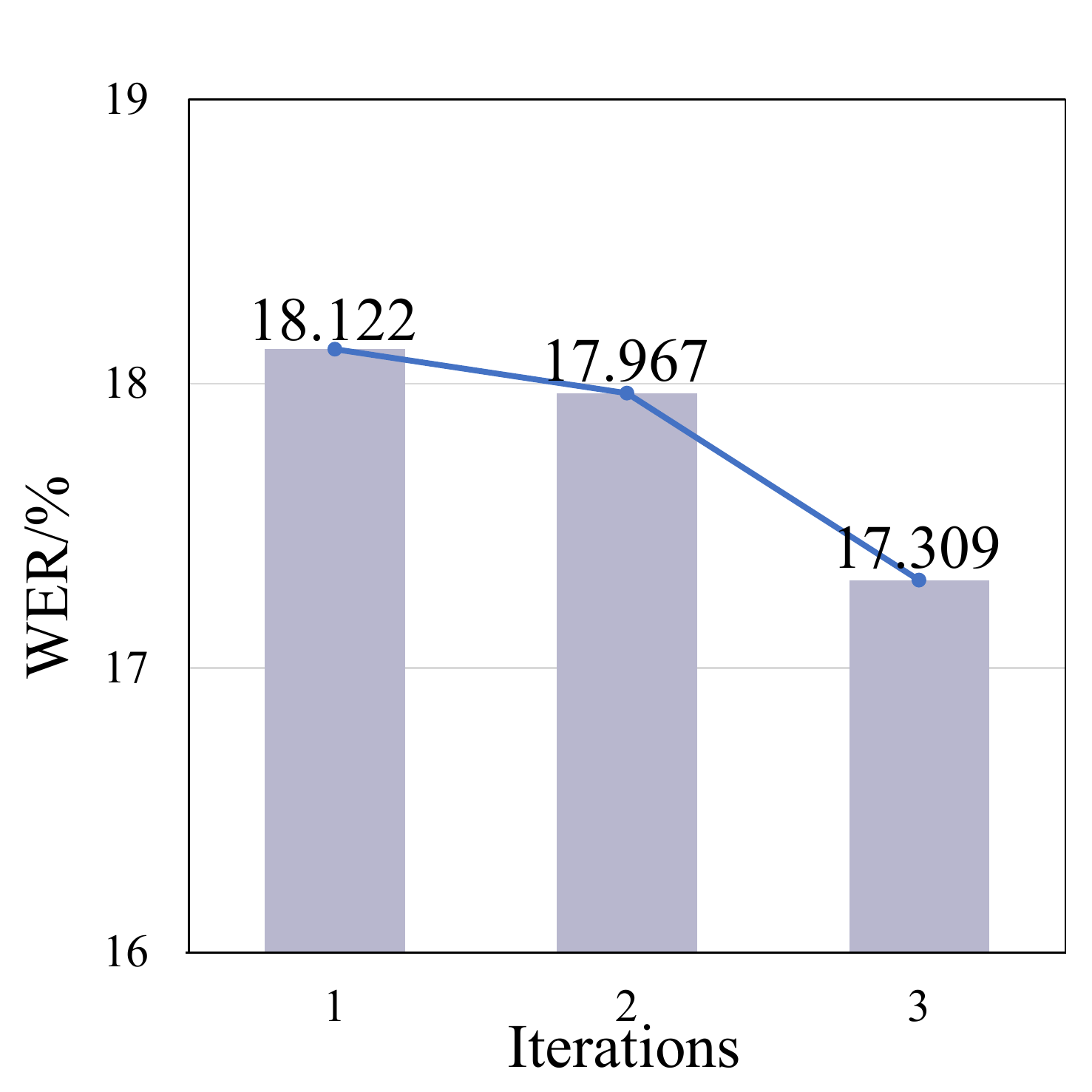}}
  \centerline{(b) Vocalized Speaking Mode}\medskip
\end{minipage}
\caption{
 The test set WERs at different training iterations of our proposed method. }
\label{WER-iters-change}
\vspace{-1mm}
\end{figure}

To examine the effectiveness of each part in our proposed method, some ablation studies were  conducted. To be concrete, we compared the performance of our proposed method, our proposed method without iterative training strategy (\textit{``w/o ITS"}), and without domain adversarial training (\textit{``w/o DAT"}) on both silent and vocalized articulation data. 
For \textit{``w/o ITS"}, the procedures of pseudo target generation and domain adversarial model training were conducted only once.
For \textit{``w/o DAT"}, Eq.~(\ref{lossoverall}) degraded to $L_{Recon,s}$ without using the domain discriminator.
As shown in Table \ref{ablation}, in both silent and vocalized speaking modes, \textit{``w/o ITS"} and \textit{``w/o DAT"} were both worse than our proposed method on all objective metrics, which demonstrates the effectiveness of our proposed iterative training strategy and domain adversarial training. 
Comparing with the results of TaLNet in Table \ref{eval-results}, we can see that both ablated models 
still outperformed TaLNet on all objective metrics. Additionally, the test set WERs at different training iterations of our proposed method were also shown in Fig.~\ref{WER-iters-change} to illustrate the effectiveness of iterative model training.

\vspace{-1mm}

\begin{figure}
        \setlength{\abovecaptionskip}{0.cm}
        \setlength{\belowcaptionskip}{-1cm}
	\centerline{\includegraphics[width=3.6cm]{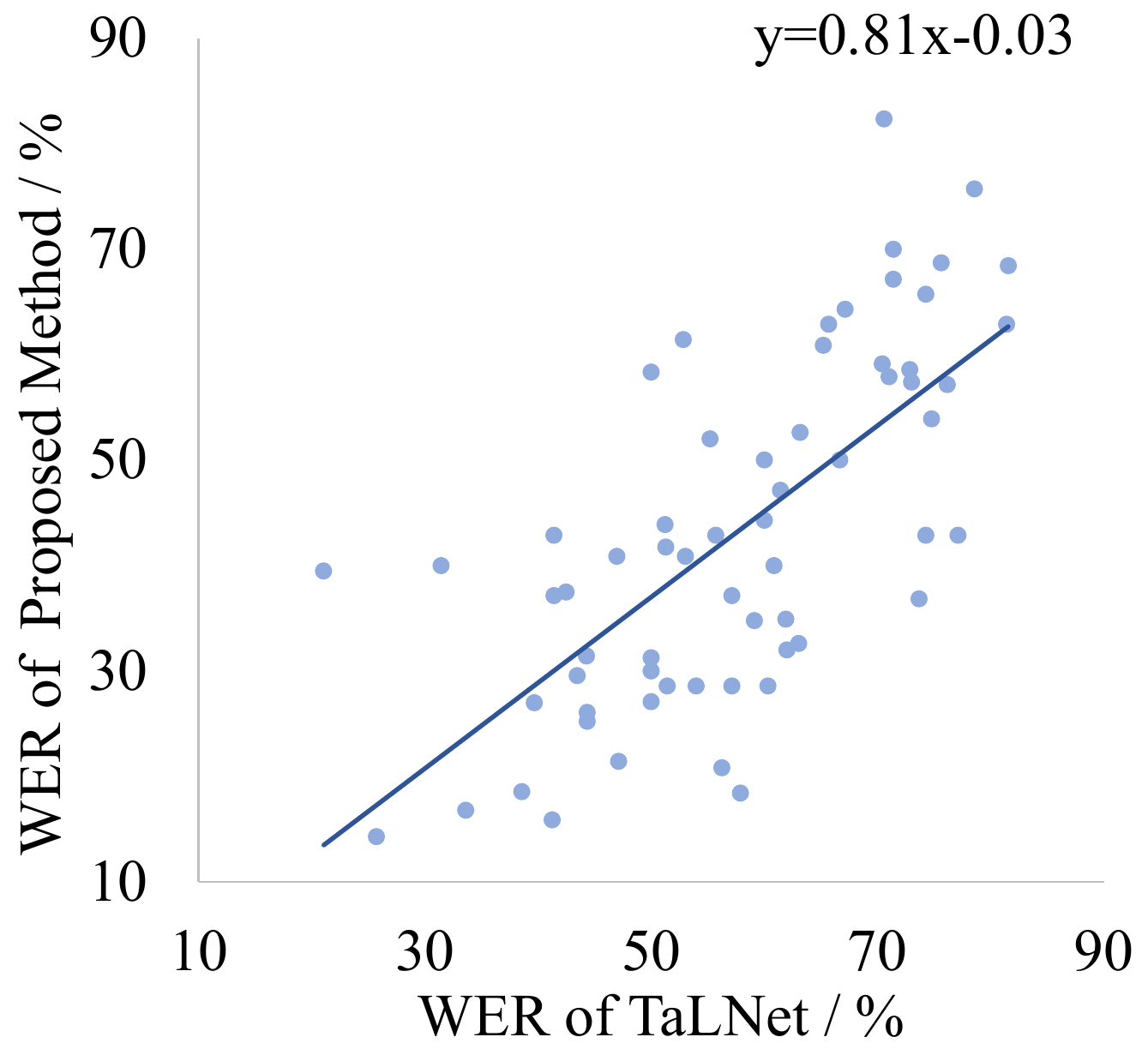}}
	%
	\caption{The LSE results of fitting the WER of speech restored using TaLNet and our proposed method. The coefficient greater than 0 indicates a positive correlation between the two methods.}
	\label{LSE}
 \vspace{-1mm}
\end{figure}

To study the performance variation among different speakers, a least squares estimation (LSE) was excuted to fit the speaker-wise WERs of speech restored using TaLNet and our proposed method. The results are displayed in Fig.~\ref{LSE}, exhibiting a strong positive correlation, i.e., 
if a speaker obtained a lower or higher WER of TaLNet while reconstructing speech in the silent speaking mode, he/she would also tend to achieve a lower or higher WER using the proposed method.
The same conclusion was also observed as in the original paper of TaLNet \cite{zhang2021talnet}, that the WER of reconstructed speech varied significantly among different speakers. 
The reason may be that 
some speakers didn't not articulate correctly when deprived of audio feedback, so it is difficult to improve the  intelligibility of their reconstructed speech using the proposed method.

\vspace{-1mm}
\section{Conclusion}
\vspace{-1mm}
 This paper has proposed using pseudo target generation and domain adversarial training to address the two major challenges in speech reconstruction from silent articualtion. Moreover, an iterative training strategy is designed to further improve the performance. Objective and subjective experimental results have demonstrated the effectiveness of the proposed method. 
 However, there is still a clear gap between the performance of articulatory-to-acoustic conversion in silent and vocalized modes. To improve the model framework by considering the intrinsic articulation differences between the two modes 
 will be the tasks of our future work.

\bibliographystyle{IEEEbib}
\bibliography{strings,refs}

\end{document}